\documentclass[aps,twocolumn,groupedaddress,pra,showpacs]{revtex4-1}

\usepackage{graphicx}
\usepackage{subfigure}
\usepackage{amsmath}
\usepackage{amssymb}
\usepackage{enumerate}
\usepackage{psfrag}
\usepackage{bm}
\usepackage{color}
\usepackage{multirow}
\usepackage[utf8]{inputenc}

\newcommand{\braket}[3]{{\ensuremath \bigl \langle \,  #1 \, \bigr | \, #2 \, \bigl | \,  #3  \, \bigr \rangle }}

\newcommand{\sket}[1]{{\ensuremath | \,  #1  \,  \rangle }}
\newcommand{\ket}[1]{{\ensuremath  \bigl | \,  #1  \, \bigr \rangle }}
\newcommand{\smallket}[1]{{\ensuremath  | \,  #1  \,  \rangle }}
\newcommand{\bra}[1]{{\ensuremath \bigl \langle \,  #1 \, \bigr | }}

\begin{document}


\title{Time-dependent restricted active space Configuration Interaction for the photoionization of many-electron atoms}

\author{David Hochstuhl}
\email[]{hochstuhl@theo-physik.uni-kiel.de}

\author{Michael Bonitz}

\affiliation{Institut f\"ur Theoretische Physik und Astrophysik, D-24098 Kiel, Germany}

\date{\today}

\begin{abstract}
We introduce the time-dependent restricted active space Configuration Interaction method to solve the time-dependent Schrödinger equation for many-electron atoms, and particularly apply it to the treatment of photoionization processes in atoms. The method is presented in a very general formulation and incorporates a wide range of commonly used approximation schemes, like the single-active electron approximation, time-dependent Configuration Interaction with single-excitations, or the time-dependent R-matrix method. We proof the applicability of the method by calculating the photoionization cross sections of Helium and Beryllium, as well as the X-ray--IR pump-probe ionization of Beryllium. 
\end{abstract}
\maketitle

\section{Introduction}
The numerical simulation of quantum mechanical many-body systems is hampered by the exponentially growing effort required to directly solve the time-dependent Schrödinger equation (TDSE). As a consequence, only one-particle systems can be described in full generality. For two-particle systems (and reduced two-particle systems)---although great progress has been made in the last decade, e.g. in the description of double ionization of Helium \cite{Foumouo_2006,Feist_2008}, molecular Hydrogen~\cite{Colgan_2002} or Beryllium~\cite{Kheifets_2001, Laulan_2004, Yip_2010}---there exist several scenarios for which direct solutions are still hardly or even not feasible: among them is, e.g., pump/probe spectroscopy~\cite{Fruehling_2009, Bauch_2010}, the laser assisted Auger decay~\cite{Schuette_2012,Bauch_2012}, and correlated high-order harmonic generation~\cite{Gordon_2006,Brown_2012}, all of which require large angular momentum expansions to obtain converged results. Further, the first-principle treatment of few- 
and many-electron systems is becoming more and more important. This can be seen, for instance, in the measurements of a time-delay in the photoionization from different atomic shells of Neon and Argon atoms~\cite{Schultze_2010}, which up to now could not been reproduced in simulations. Other important processes which would profit from direct solutions, beside those already mentioned, are tunnel ionization~\cite{Brabec_2005} and excitation of hole states~\cite{Pabst_2011,Kuleff_2011}. In this work, we introduce a method to the field of numerical simulations of photoionization, which is capable of treating a broad class of time-dependent physical processes, including the ones mentioned, in a very general manner.

Several methods have been designed to attack the many-particle TDSE by using a reduced description, such as, e.g., time-dependent density functional theory~\cite{Chu_2001}, Nonequilibrium Green functions~\cite{Hochstuhl_2010_Green,Balzer_2010,Bonitz_2010} or semi-empirical approaches~\cite{Corkum_1993,Santra_2006}; they often are applicable to rather large systems, but typically lack a control of the accuracy. Regarding the wavefunction based schemes, we mention the Multiconfigurational time-dependent Hartree-Fock (MCTDHF) method~\cite{Meyer_1990,Hochstuhl_2010,Bonitz_2010,Hochstuhl_2011,Haxton_2011}, the time-dependent Configuration Interaction singles method (TD-CIS)~\cite{Greenman_2010} and the single- and two-active electron approximation (SAE/TAE)~\cite{Schafer_1993,Kamta_2002}. Further, there is the time-dependent R-matrix method (TD-RM)~\cite{Lysaght_2008,Guan_2009}, which can be considered the most successful time-dependent approach to few-electron systems so far.
The time-dependent restricted active space Configuration Interaction (TD-RASCI) method employed in this work can be considered as a superset to all the mentioned determinant schemes: it contains the TD-CIS, SAE/TAE methods, and even TD-RM as special cases, and it can also be used to extend the range of the MCTDHF method. As will be discussed later, due to its generality, we believe that it should be applicable to processes which up to now could not be efficiently described in direct calculations. The restricted active space idea is well known for over twenty years in quantum chemistry~\cite{Olsen_1988}. The goal of this paper is the extension to the time-dependent regime and the optimization for photoionization processes. 

In this work, we aim at presenting the first test calculations, without fully exploiting the capabilities of the TD-RASCI method. Rather, as a first step, we focus on a standard problem in photoionization, namely the calculation of total cross sections of the atoms Helium and Beryllium. We therefore employ an explicitly time-dependent description, and compare different approximations for the wavefunction with experimental and theoretical results from the literature. Our results show that the TD-RASCI method yields an accurate description of the doubly-excited states at a fraction of the effort of full direct solutions. As a further example, we consider the XUV-IR pump-probe process in Beryllium.

The paper is organized as follows: section II introduces the full and restricted configuration interaction schemes, which are subsequently related to commonly used approximations in the treatment of photoionization, and gives the main ideas of our numerical implementation. In section III, we particularly focus on the treatment of photoionization processes. Therefore, we introduce a partitioning of the coordinate space, as well as a mixed single-particle basis, which will turn out a very convenient ingredient of the present method. Section IV presents the first numerical results for Helium and Beryllium. In section V, we summarize the concepts and results.

\section{Time-dependent Restricted Active Space Configuration Interaction (TD-RASCI)}\label{sec:td_rasci}
\subsection{Full Configuration Interaction}
Our aim is to solve the time-dependent Schrödinger equation (TDSE)
\begin{align}
\label{eq:Schroedinger_equation} i \partial_t \, \ket{\Psi(t)} \ = \ \hat H(t) \; \ket{\Psi(t)}\,,
\end{align}
for the $N$-particle wavefunction $\sket{\Psi(t)}$, which provides the complete information of the system (in a pure state description). 
We concentrate on a Hamiltonian describing Coulomb-interacting fermions in an atom of charge $Z$ subjected to an external laser field (we use atomic units),
\begin{align}
\label{eq:Hamiltonian} \hat H(t) \ = \ \sum_{k=1}^N \, \Biggl\{ \frac{{\mathbf {\hat p}_k}^2}2 -  \frac{Z}{r_k}  +  \mathbf E(t) \,\mathbf r_k \Biggr\}  \, + \,  \frac12  \, \sum_{k\neq l}  \frac{1}{|\mathbf r_k-\mathbf r_l|}\,.
\end{align}
The method presented in the following, however, is completely general and might be applied to a large variety of other physical systems. We begin with the expansion of the wavefunction in a set of Slater determinants,
\begin{align}
\label{eq:wavefunction_expansion} \ket{\Psi(t)} \ = \ \sum_{I\in\Omega} C_I(t) \ \ket{\psi_{i_1} \psi_{i_2} \cdots \psi_{i_N}}\,,
\end{align}
in which the multi-index $I=(i_1,\cdots,i_N) \in \Omega$ specifies the occupied single-particle spin-orbitals $\smallket{\psi_k}$, which are assumed to be orthonormal throughout. The sum is performed over an index-set $\hspace{0.04em} \Omega \subset \mathbb N^N$, which determines the set of Slater determinants included in the expansion, and thus the accessible subspace of the Hilbert space $\mathcal H_N$ in which the wavefunction lives. By insertion of the ansatz (\ref{eq:wavefunction_expansion}) into the TDSE, we obtain the equation of motion for the expansion coefficients,
\begin{align}
\label{eq:Schroedinger_equation_discretized} i \, {\dot C}_I(t) \ = \ \sum_{J\in\Omega} \braket{I}{\hat H(t)}{J} \; C_J(t)\,,
\end{align}
which is just the TDSE projected onto the subspace of $\mathcal H_N$ defined by $\Omega$. In order to solve this equation, the arising Hamiltonian matrix elements have to be evaluated using Slater-Condon rules, see e.g. Ref.~\cite{Helgaker}.

To proceed, one needs to specify the index set $\Omega$. By choosing
\begin{align}
\Omega_\text{FCI} \; = \; \big\{(i_1,\cdots,i_N)\ \big| \ 1\leq i_1 < \cdots < i_N \leq 2N_b \big\} \,,
\end{align}
one obtains the Full Configuration Interaction (FCI) ansatz. The determinant basis has a size of $\binom{2N_b}{N}$ and spans the maximal accessible Hilbert space for a given spin-orbital basis of size $2N_b$. The corresponding wavefunction and operators are, up to the discretization and errors in the time-integration, represented exactly. In particular, the wavefunction contains the entire possible correlation, and its solution provides the benchmark result for any approximate method in the same single-particle basis. In the limit of a sufficiently accurate single-particle basis, one essentially recovers the true result. 

Eqs.~(\ref{eq:wavefunction_expansion}) and (\ref{eq:Schroedinger_equation_discretized}) present the basis of all direct approaches, though they are often stated in a different form. In the photoionization community, and particularly in the treatment of two-particle systems, one often encounters the close-coupling ansatz, where the expansion is made in angular-momentum eigenfunctions. In quantum chemistry, on the other hand, it is common to work in a basis of spin-eigenfunctions, so called configuration state functions~\cite{Pauncz}. Both types of basis sets may be obtained from the Slater-determinant basis by taking appropriate linear combinations, the expansion coefficients of which are given in terms of Clebsch-Gordan coefficients. In a symmetry adapted basis set, one typically has to deal with less basis states and also obtains a clearer interpretation of the wavefunction, but this often goes at the cost of a more difficult evaluation of Hamiltonian matrix elements (particularly for more than two 
particles).

As mentioned previously, the time-dependent Full Configuration Interaction method can hardly be applied to photoionization processes of $N>2$ particle systems (in three dimensions), since the modeling of an adequate continuum often requires a rather large single-particle basis. This causes the FCI expansion to become unfeasibly large. A trivial solution is to employ smaller basis sets resp. grids, as it is done in quantum chemistry calculations or in a recent time-dependent close-coupling study of the Lithium atom~\cite{Colgan_2012}. The RAS method presented in the next section takes another approach: it retains the accuracy of the single-particle basis, but restricts the wavefunction on the many-body level.

\subsection{Restricted active space Configuration Interaction}
Time-dependent restricted active space Configuration Interaction (TD-RASCI) provides a way to effectively reduce the determinant basis size. The underlying idea is quite simple: given the huge Full-CI space, remove all parts of this space which expectedly will not be occupied by the wave function. Practically, this means that a certain set of Slater determinants is dropped from the expansion (\ref{eq:wavefunction_expansion}), which reduces the size of the discretized Hilbert space and thus facilitates the numerical solution. The RAS method formalizes this idea and provides a systematic approach for the selection of the important determinants. 
It is routinely used in quantum chemistry for more than 20 years~\cite{Olsen_1988}, and there especially in Configuration Interaction and Multiconfigurational Hartree-Fock calculations. In this context, it is also termed restricted active space self-consistent field (RASSCF). The idea of selecting individual configurations, however, has already been used much earlier and was applied, e.g., in Configuration Interaction singles- and/or doubles calculations~\cite{Shavitt_1998}.
In the time-dependent treatment of photoionization processes, up to now, we are only aware of special cases, such as the single-active electron approximation (SAE) or time-dependent Configuration Interaction singles (TD-CIS). They will be related to the present method later in this work.

\begin{figure}[!t]
 \begin{center}
   \includegraphics[width=0.8\linewidth]{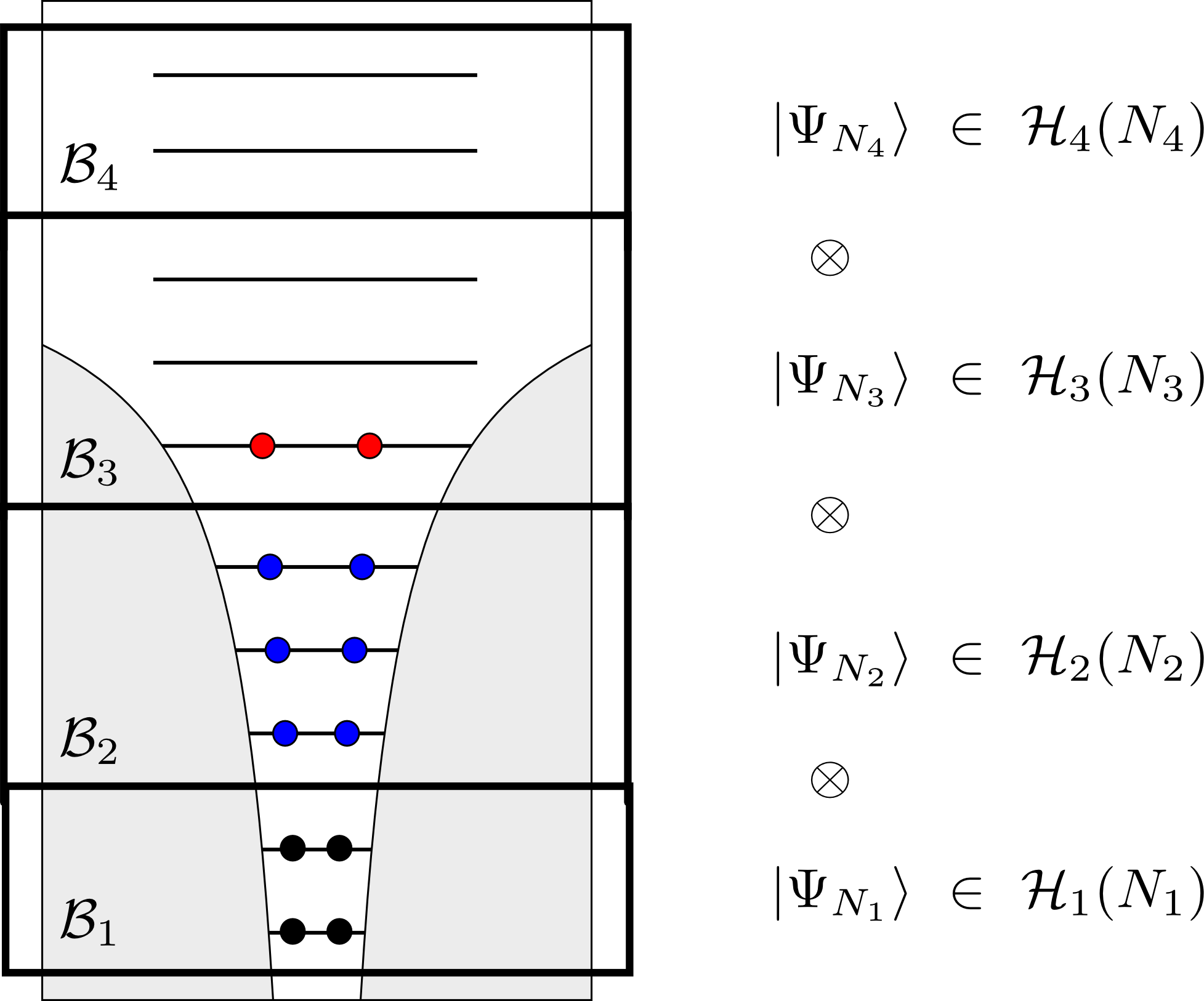}
 \end{center}
 \caption{\label{fig:RASCI} Illustration of the restricted active space scheme for a number of $P=4$ partitions of the single-particle basis $\mathcal B$. In each partition, one imposes certain restrictions on the allowed particle numbers. The total $N$-particle wavefunction $\sket{\Psi}$ is constructed as the tensor product of the $N_i$-particle wavefunctions $\sket{\Psi_i}$, with $\sum_i N_i = N$. }
\end{figure}

The RAS method is formally obtained by imposing restrictions on the set $\Omega$ of allowed Slater determinant indices. This, however, can be cumbersome without a clear and intuitive picture in mind. Therefore, a useful procedure is given by the following steps:
\begin{enumerate}[(i)]
\item Divide the single-particle basis
\begin{align}
\mathcal B \ = \ \big\{\ket{\psi_1},\cdots, \ket{\psi_{2N_b}}\big\}
\end{align}
into an arbitrary number $P$ of partitions $\mathcal B_i$,
\begin{align}
\label{eq:single_particle_basis_RAS}\mathcal B \ =\ \mathcal B_1 \, \cup \, \cdots \, \cup \,  \mathcal  B_{P}\,,
\end{align}
with
\begin{align}
\mathcal B_i\ = \ \big\{\ket{\psi_{p_{(i-1)}}},\cdots, \ket{\psi_{p_{i}-1}}\big\} \,.
\end{align}
A division is thus defined by the $P+1$ numbers $(p_0=1,\,p_1, \cdots , p_{P-1},\,p_P=2N_b)$, i.e. by $P-1$ free parameters. The number of orbitals in the partition $\mathcal B_i$ is denoted by $N_{b,i}$; by definition, one has $\sum_j N_{b,j} = 2N_b$. The partitions are visualized by the four black boxes in Fig.~\ref{fig:RASCI}.

\item Impose restrictions on the allowed particle numbers. Therefore, for each partition $\mathcal B_i$, we specify the minimal and maximal particle number, $N_{\text{min},i}$ and $N_{\text{max},i}$, and allow only for particle numbers $N_i$ in between these two values, $N_{\text{min},i}\leq N_i \leq N_{\text{max},i}$. The restrictions should be assigned to match the occurring physical processes as good as possible, but at the same time result in only a moderate number of determinants.
\end{enumerate}

Each pair $(\mathcal B_i, N_i)$ obtained this way is related to a discrete Hilbert space $\mathcal H_i(N_i)$, which is the span of all $N_i$-particle Slater determinants constructed from the truncated single-particle basis $\mathcal B_i$. For an example, see the right-hand part of Fig.~\ref{fig:RASCI}. The total Hilbert space $\mathcal H_\text{RAS}$ is thus decomposed as
\begin{align}
\mathcal H_\text{RAS} \ = \hspace{-1.5em}\bigcup_{\substack{\\[0.25em] N_{\text{min},i}\leq N_i \leq N_{\text{max},i} \\[0.3em] \sum_j N_j = N}} \hspace{-1.5em} \mathcal H_1(N_1) \times \cdots \times \mathcal H_P(N_P)\,,
\end{align}
i.e. as the unification of the Cartesian products of all sub-Hilbert spaces which have the correct particle number $N$ and satisfy the RAS constraints. It has the total dimension
\begin{align}
\dim(\mathcal H_\text{RAS}) \ = \hspace{-2em}\sum_{\substack{\\[0.6em] N_{\text{min},i}\leq N_i \leq N_{\text{max},i} \\[0.3em] \sum_j N_j = N}} \hspace{-1.5em} \binom{N_{b,1}}{N_1} \cdots \binom{N_{b,P}}{N_P}\,,
\end{align}
and the wavefunction is set up as the antisymmetrized product of the given Slater determinants (which is just a determinant in the original Hilbert space),
\begin{align}
\ket{\Psi(t)} \ = \hspace{-0.8em}\sum_{\substack{\\[0.3em] \sket{I_i} \in \mathcal H_i(N_i)}} \hspace{-0.8em} C_{I_1,\cdots,I_P}(t) \ \hat{\mathcal A} \; \ket{I_1} \cdots \ket{I_P}\,.
\end{align}
Finally, in order to bring the wavefunction expansion to the form of Eq.~(\ref{eq:wavefunction_expansion}), we need to specify the set of allowed Slater determinant indices, 
\begin{align}
\Omega_\text{RAS} =  \big\{(i_1,\cdots,i_N) \in \Omega_\text{FCI}\ \big| &\ \forall j   \in \{1,\cdots,P\} \, : \\\notag & N_{\text{min},j} \leq N_j \leq N_{\text{min},j}\, \big\} \,,
\end{align}
Here, the occupation $N_j$ of the $j-$th partition $\mathcal B_j$ is given by
\begin{align}
N_j(I;p_1,\cdots,p_{P-1}) \ = \ \sum_{k=1}^N \left\{ \begin{array}{ll} 1\,, \quad& p_{j-1}\leq i_k < p_j \\[0.3em] 0\,, & \text{else}\end{array} \right\}\,.
\end{align}
Note that the notation $\Omega_\text{RAS}$ does not explicitly state the strong dependence on the RAS parameters, i.e. on the partitioning $(p_1,\cdots,p_{P-1})$ and the range of allowed particle numbers defined by $N_{\text{min},i}$ and $N_{\text{max},i}$.

If we would have made no restriction on the particle numbers, i.e. allowed for $0 \leq N_i \leq N$ in each partition $\mathcal B_i$, we would essentially recover the Full Configuration Interaction method. By restricting the accessible many-body Hilbert space in the way just presented, one can significantly reduce its dimension and thus enable time-dependent Configuration Interaction calculations which are far beyond reach of the Full CI scheme.

\begin{figure}[!t]
 \begin{center}
   \includegraphics[width=0.95\linewidth]{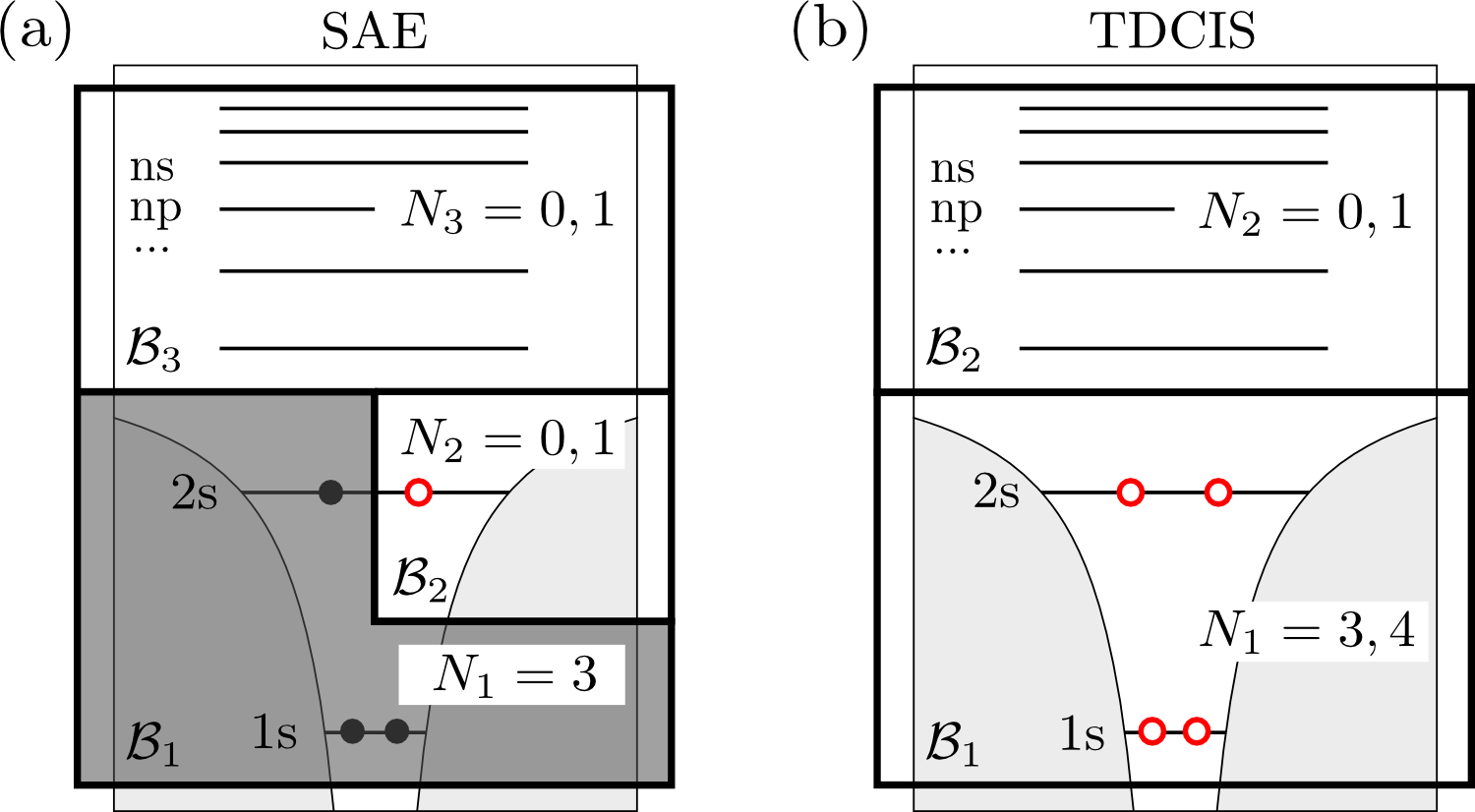}
 \end{center}
 \caption{\label{fig:SAE_TD-CIS} Special cases of the TD-RASCI scheme for the example of Beryllium. The numbers $N_i$ label the allowed particle numbers in the partitions $\mathcal B_i$ (black boxes). Active electrons are shown as open (red) circles. \emph{Left}: Single-active electron approximation with an active $2s$ orbital. The gray-shaded orbitals are fixed, and no more than a single electron is allowed in the continuum. \emph{Right}: Time-dependent Configuration Interactions singles. All electrons are active, but only single excitations from the groundstate are included.}
\end{figure}

\subsection{RAS examples: SAE, TAE, TD-CIS}
The previous section might be regarded a bit technical, as we considered the formalism in a very general fashion. Actual applications of the RAS scheme are often much simpler. To get a feeling for its capabilities, the present section shall give the first examples where the RAS scheme reduces to (variants of) some well-known methods in the treatment of photoionization. At the same time, we wish to show how natural one can arrive at an extension of these standard methods.
Let us start with the single-active electron (SAE) approximation. There, the $N$-particle problem is reduced to a single-particle problem by freezing all electrons except the active one. An example of a corresponding basis partitioning for Beryllium is shown in the left part of Fig.~\ref{fig:SAE_TD-CIS}. The total number of Slater determinants is given by $2N_b-N$, i.e. one determinant for each spin-orbital the active electron can reach. If we further include the spin-symmetry requirement $\langle S_z \rangle =0 $, the number of determinants reduces to $N_b-N$, as then only the spin-up orbitals are accessible. Similarly, also $\langle L_z\rangle=0$ could be imposed. With the effort scaling only linearly with $N_b$, the solution of the corresponding equation of motion~(\ref{eq:Schroedinger_equation_discretized}) has only the complexity of a single-particle problem.

In the particular form commonly applied in strong-field physics, the SAE approximation has been criticized for not being an \emph{ab-initio} method~\cite{Rohringer_2006}. This is due to the reduction of an $N$-particle problem to merely a one-particle equation, and the related need to estimate the effective potential experienced by the active electron. We want to stress, that this does not hold for the SAE approximation presented here, as the whole derivation assumed throughout an $N$-particle representation.
Similar variants of this extended SAE approximation have already been given in Refs.~\cite{Rohringer_2006,Spanner_2009}. Moreover, the scheme is easily extendable: in the case that another electron is likely to get ionized, one could treat the related orbital active as well, and thereby obtain the two-active electron (TAE) approximation. The TAE approximation has been applied, for instance, to Beryllium~\cite{Kamta_2002,Laulan_2004,Yip_2010}. In these works, however, like in the SAE approximation one has to assume effective potentials, which are often of Hartree-Fock type and held fixed during the propagation. In contrast, the improved TAE approximation within the RASCI scheme contains the exact interaction terms, which are consistently adjusted during the propagation. The effective two-particle problem obtained in the TAE approximation can already be difficult to solve for large basis sizes $N_b$. Therefore, in order to retain the single-particle complexity, one could further restrict the number of allowed 
electrons in the virtual orbital space to one, which would lead to an only twice as large problem compared to a single active electron.

Repeated application of this idea leads to the time-dependent Configuration Interaction singles (TD-CIS) approximation, where each of the $N$ electrons is considered as active, but only a single particle is allowed in the virtual orbital space at the same time. The visualization of this constraint is shown in Fig~\ref{fig:SAE_TD-CIS}.
The TD-CIS scheme has been applied with success to processes such as high-order harmonic generation in noble gas atoms~\cite{Gordon_2006} or hole excitation processes~\cite{Pabst_2011}. It naturally lacks, however, the description of transitions leading to doubly and higher excited states. As will be shown later in the results section, the RAS scheme makes it easy to add certain selected states, and by this can account e.g. for the doubly-excited resonances in Helium. The TD-CIS approximation could also be extended to include all double (TD-CISD) and higher (TD-CISDT, etc) excitations, however, at the price of a significantly increased effort.

\subsection{Numerical implementation}
In the following we summarize the main ideas of our numerical implementation of the TD-RASCI method. One of its great advantages is that, when properly implemented, it comprises a lot of different approximations in a single and generic program, which are achieved by a simple change of the RAS parameters.
Restricted active space CI follows the basic work cycle as found in most CI implementations. In particular, the steps to accomplish are
\begin{enumerate}[(1)]
 \item Choose the number of particles $N$, and an appropriate single-particle basis $\mathcal B=\{\sket{\psi_k}\}$.
 \item Select the RAS constraints, i.e. a partitioning of the basis into $P$ elements $\mathcal B_i$ and corresponding restrictions on the particle numbers $N_i$.
 \item Construct the determinant basis and calculate the Hamiltonian matrix using the Slater-Condon rules.
 \item Solve the time-independent Schrödinger equation to obtain the initial eigenstate.
 \item Integrate the time-dependent Schrödinger equation up to the required system time $T$.
\end{enumerate}

The first step is similar to Full CI calculations, although some care has to be devoted to the selection of the single-particle basis. Our choice is described in detail below, in section~\ref{sec:basis_set}; essentially, the basis should be flexible enough to describe each possible state the electrons may occupy. The fourth and fifth step, the numerical solution of the Schrödinger equation, is a standard task that can be accomplished using solvers like the Lanczos or Davidson method in the time-independent case~\cite{Saad_1992}, and propagators like the short iterative Lanczos~\cite{Park_1986} or the Crank-Nicolson~\cite{vanDijk_2007} method for the time-dependent version. For efficiency, they should be implemented using sparse matrix algebra. Note, however, that depending on the RAS constraints, the degree of sparsity may be smaller than in the FCI case.

A true modification with respect to Full CI is made only in the second step, since RAS schemes require a more elaborate ordering and bookkeeping of the determinants. In order to construct the Hamiltonian matrix for the system (\ref{eq:Hamiltonian}), one needs to be able to calculate the matrix elements
with Slater determinants $\sket{I}$,$\sket{J}$
\begin{align}
\notag H_{IJ}(t) \ &= \ \braket{I}{\hat H(t)}{J} \ = \ \sum_{pq} h_{pq}(t) \braket{I}{\hat a^\dagger_p \hat a_q}{J} \\
\label{eq:Hamiltonian_second_quant}&\quad + \frac 12 \sum_{pqrs} g_{pqrs} \braket{I}{\hat a^\dagger_p \hat a^\dagger_r \hat a_s \hat a_q}{J}\,.
\end{align}
In the last equality, we inserted the second quantization representation of the Hamiltonian, see e.g. Refs.~\cite{Hochstuhl_2011,Helgaker}, which is given here as the sum of an explicitly time-dependent single-electron part (first sum containing kinetic and potential energy and the external field) plus a two-particle part (the second with the Coulomb interaction). The operators $\hat a_p$ ($\hat a^\dagger_p$) thereby annihilate (create) a particle in the spin-orbital $\sket{\psi_p}$. The evaluation of the total Hamiltonian is sketched in the following by means of the single-particle term. For a more detailed introduction, we refer to Ref.~\cite{Helgaker}. First, one loops over the set of determinants $\sket{I}$ and the spin-orbital indices $(p,q)$ [resp. $(p,q,r,s)$] and applies the excitation operators e.g. to the left, to obtain
\begin{align}
\bra{I}\; \hat a^\dagger_p \hat a_q \ = \ \xi \,\bra{J}\,,
\end{align}
with the phase factor $\xi \in \{-1,0,1\}$. Note that one only gets a non-zero contribution if orbital $p$ is occupied in $\sket{I}$, and either orbital $q$ is unoccupied or $p=q$ holds. Next, the term $\xi \, h_{pq}$ should be added to the Hamiltonian matrix element $H_{IJ}$. One therefore needs an efficient scheme to retrieve the address $\text{add}(\sket{J})$ of the determinant $\sket{J}$. The simple idea of a linear search in the determinant list is thereby inappropriate, as that would imply a quadratic scaling in the number of determinants. A more convenient method is to construct the addresses directly from the determinant by using graphical or combinatorial techniques~\cite{Helgaker,Klene_2000,Klene_2003,Streltsov_2010}. Following Ref.~\cite{Knowles_1984}, in the Full CI case with $M$ orbitals and $N$ particles, one obtains
\begin{align}
\text{add}(\sket{I}) \ = \ 1+ \sum_{p=1}^N Z(p,i_p)\,,
\end{align}
with the function $Z$ defined as
\begin{align}
Z(k,l) \ = \ \sum_{m=M-l+1}^{M-k} \left[\binom{m}{N-k}-\binom{m-1}{N-k-1} \right]\,,
\end{align}
for $k<N$ and $Z(N,l) \ = \ l-N$.
With this choice, the address function $\text{add}(\sket{I})$ assigns a one-to-one mapping from the set of $N_\text{det}=\binom{M}{N}$ Slater determinants $\sket{I}$ to the set of indices $\{1,\cdots,N_\text{det}\}$.

As introduced above, RAS Hilbert spaces are composed as tensor products of a certain number of smaller FCI Hilbert spaces, and RAS determinants are given as tensor products of FCI determinants. It is thus not too difficult to evaluate the address in the FCI subspaces, and connect these values to obtain the address in the total RAS space. For a detailed derivation, see e.g. Refs.~\cite{Klene_2003,Mendl_2011}.

\section{Application to photoionization}

\subsection{Spatial partitioning}
\begin{figure}[!t]
 \begin{center}
   \includegraphics[width=0.85\linewidth]{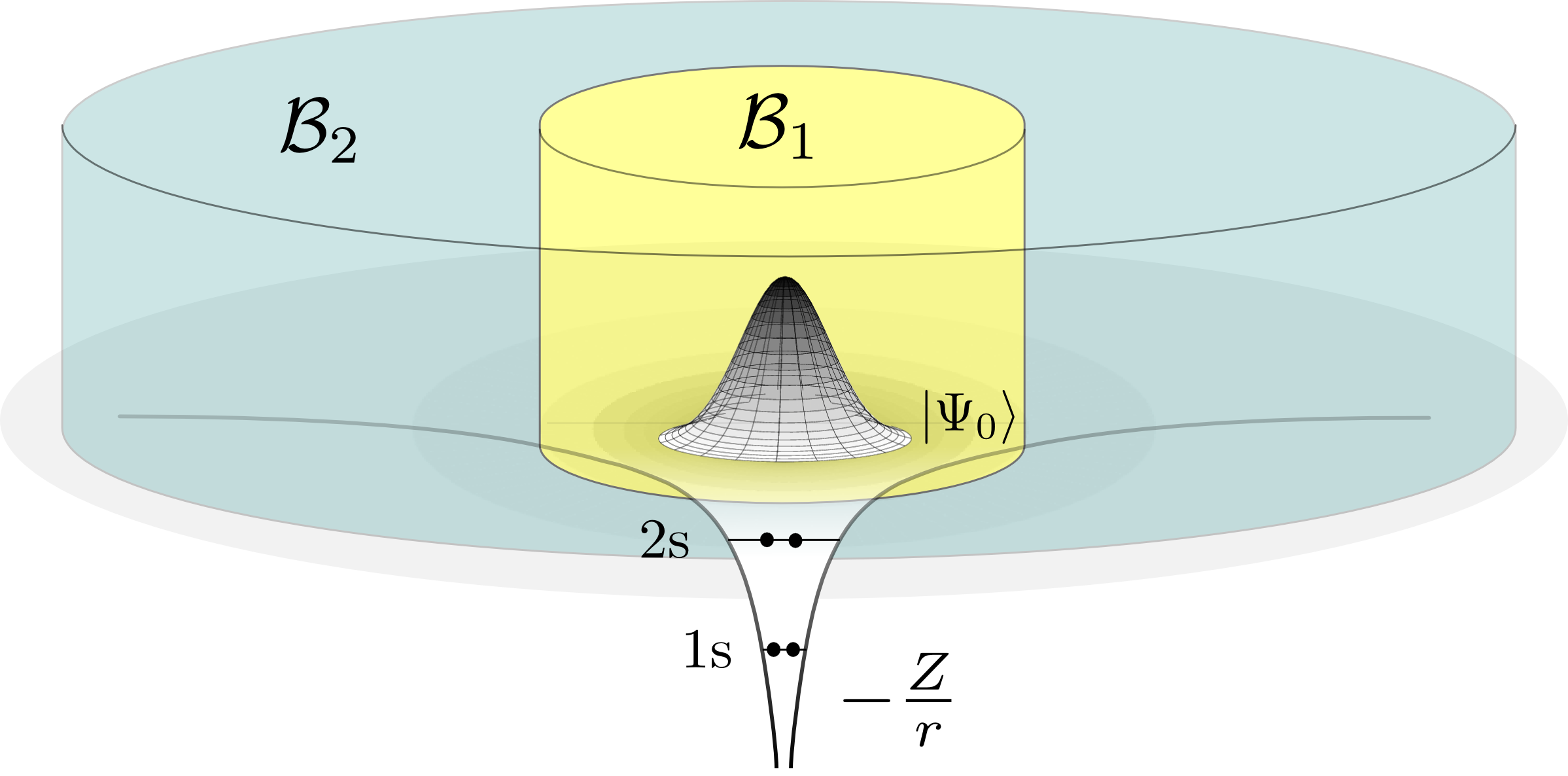}
 \end{center}
 \caption{\label{fig:RASspatial}Spatial partitioning of the active space used in the RASCI treatment of photoionization processes. $\mathcal B_1$ marks the region in which the groundstate $\sket{\Psi_0}$ is localized, $\mathcal B_2$ the continuum. The example shows a four electron atom (e.g., beryllium).}
\end{figure}

The RAS schemes applied in quantum chemistry are commonly considered in energy space, that is, the RAS constraints are chosen according to the orbital energies. Typically, a penalty is placed on the high-lying energy orbitals, since these are expected to contribute only little to the targeted groundstate or low excited states. This view is also indicated in Figs.~\ref{fig:RASCI} and \ref{fig:SAE_TD-CIS}. For an efficient treatment of photoionization processes, it is crucial to extend this concept also to the coordinate space, where the partitioning is, instead, performed according to the spatial regions. A basic example is shown in Fig.~\ref{fig:RASspatial}. There, the coordinate space is divided into a region in the vicinity of the nucleus ($\mathcal B_1$) and a region outside ($\mathcal B_2$). Over the region near the atomic core we construct the Hilbert space $\mathcal H_1$, which should appropriately describe the groundstate~$\sket{\Psi_0}$, while the Hilbert space $\mathcal H_2$ constructed over the 
outside region is used to model the scattering states. 
In the same way as before, one can then restrict the allowed particle numbers; for instance, one could consider only single-ionization processes, and thus allow only for a single electron in $\mathcal B_2$. Note that here one implicitly makes the assumption that the true singly-ionized states are reasonably well described by singly excited determinants. In order to perform the spatial partitioning, a single-particle basis is required which is localized in the partitioned regions, as described in the next section.

The introduced ``division of space''-concept is traditionally employed in the R-matrix method, which currently presents the most successful approach to the photoionization problem of many-electron atoms. R-matrix calculations are usually performed in a time-independent formulation, e.g. by using Floquet theory~\cite{Burke_1991}, though recently also a time-dependent versions have been proposed~\cite{Lysaght_2008,Guan_2007,Guan_2008}. Shortly, the R-matrix method proceeds in the following way: After the division into an inner and an outer region, the wavefunction in the inner region is expanded in a set of angular-momentum eigenfunctions. The focus is mostly on single-ionization, i.e. only one electron may traverse to the outer region, where it is described, e.g., on a finite-difference grid~\cite{Lysaght_2011} or through B-splines~\cite{Guan_2007,Guan_2008}; occasionally, also double ionization is considered~\cite{Guan_2009}. The interaction of the outer electron with the inner electrons is described by 
using a multipole expansion, while the effect of the outer on the inner electrons is neglected. At the boundary, a set of equations has to be solved in order to connect the wavefunctions in the two regions, which are known as the R-matrix equations. Hence, the R-matrix method takes a different strategy than the RAS approach: while the R-matrix method starts from two separated spaces and connects them by a fitting procedure, the RAS scheme begins with a large connected Full CI space and retains only a selected subspace.

In the formulation as stated before, the R-matrix method can be considered as a special case of the TD-RASCI method presented in this work (one could, however, also think of more general expansions in R-matrix theory). In order to arrive there, method one needs to apply the same spatial partitioning in the TD-RASCI and allow only for single electrons in the outer region, while the wavefunction in the inner region needs to be expressed in angular-momentum eigenfunctions. Next, one has to approximate the electronic Coulomb interaction accordingly, i.e. apply a multipole expansion to certain contributions resp. neglect other contributions. This would surely considerably speedup of the calculation. We note, however, that in its natural formulation the RAS scheme treats all the arising interactions in an exact way, i.e. without any multipole expansion.
Moreover, the RAS scheme offers several possibilities for extensions, which are simply achieved by changing the input parameters. For example, it may be easily extended to
\begin{enumerate}[(i)]
\item the treatment of double-ionization. This is easily accomplished by allowing for two electrons in the outer region.
\item use more spatial subdivisions. For example, one could allow for double ionization only in a small region and for single ionization in a much larger region.
\item apply a combined spatial and energy partitioning. For instance, the double-ionization could be restricted to states with angular momentum $l=1$, while in the other continuum states only single-ionization is allowed. This can be useful, e.g., in the treatment of laser-assisted Auger decay.
\end{enumerate}
Note that for each of the mentioned scenarios, the equation of motion is simply given by the discretized TDSE~(\ref{eq:Schroedinger_equation_discretized}), and it is not necessary to derive specialized equations. The only task is to adjust the RAS restrictions once in the beginning to the physical problem at hand.

The R-matrix method had great success in the description photoionization processes and provided several benchmark results for many-electron atoms.
This makes us confident that the RAS scheme may be suitable as well, and that it hopefully can extend the range of applications to scenarios which could not be treated efficiently so far.

\subsection{Spherical FEDVR basis}

As mentioned in the previous section, to apply the spatial partitioning the single-particle orbitals should be localized in the spatial partitions. For the present treatment of photoionization, the basis should further be sufficiently flexible to describe the scattering states.
These two requirements rule out several commonly used basis sets; for instance, Slater- and Gaussian-type functions, which are routinely used in quantum chemistry (and occasionally also in photoionization studies~\cite{Krause_2005}), are not able to provide an adequate continuum. Other basis sets like Sturmian functions, though useful for photoionization~\cite{Foumouo_2006}, are not spatially localized. Here, we apply a partial wave expansion~\cite{Hochstuhl_2011},
\begin{align}
\label{eq:single_particle_basis} \psi_{klm}(\mathbf r,m_S) \ = \ \frac{\chi_k(r)}{r} \; Y_{lm}(\theta,\phi)\, \sigma(m_S)\,,
\end{align}
where $Y_{lm}$ denotes a spherical harmonic, and $\sigma$ a spin-eigenfunction ($\sigma\in\{\alpha,\beta\}$).
In order to be localized, the radial basisfunctions should be represented by finite-difference grids, B-splines or discrete variable representations. Note that this choice implies that we can construct the partitions only with respect to the radial distance to the nucleus, but not to the angles. The use of orbitals of this kind further ensures that the Slater determinants are eigenfunctions of the orbital angular-momentum operator $\hat L_z$ and the spin-projection $\hat S_z$.

In this work, we use the finite-element discrete variable representation (FEDVR) basis for the radial basisfunctions $\chi(r)$~\cite{Rescigno_2000}, which is illustrated in Fig.~(\ref{fig:mixed_basis}). The coordinate space is thereby divided into a chosen number of finite elements, and in each element the basisfunctions are given by (normalized) Legendre interpolating polynomials,
\begin{align}
\label{eq:FEDVR_basis} {\chi_k(r)} \ = \ \frac{1}{\sqrt{w_k}} \; \prod_{j\neq k} \frac{r-r_j}{r_k-r_j} \,,
\end{align}
which are constructed over a Gauss-Lobatto grid $\{r_k\}$ with integration weights $\{w_k\}$. Additionally, at the boundary between two finite elements a bridge function is introduced to ensure the continuity of the wavefunction. For a detailed construction of the FEDVR we refer to Refs.~\cite{Rescigno_2000,Schneider_2006}. For our purpose, the use of a FEDVR basis is of great convenience due to several reasons: first, the interpolating polynomials~(\ref{eq:FEDVR_basis}) are sufficiently flexible to represent very general functions. Next, the matrix representation of local operators and the kinetic energy is sparse, which in turn leads to a sparse representation of the many-body Hamiltonian. And third, the finite-elements can be easily adjusted to model the chosen spatial division.

\subsection{Mixed basis\label{sec:basis_set}}
In order to obtain a well prepared groundstate in the RAS calculations, one needs to perform a further adjustment of the single-particle basis introduced before. Let us first shortly sketch the problem: Full Configuration Interaction Hilbert spaces are invariant under unitary rotations of the single particle basis. That is, the Hilbert spaces spanned by all $N$-particle determinants constructed from a given single-particle basis $\{\sket{\psi_k}\}$ of size $N_b$ is exactly the same as if we would use any unitarily transformed basis $\sket{\tilde \psi_j} = \sum_j U_{jk} \sket{\psi_k}$. While being mathematically equivalent, different choices in general differ in their numerical behavior, and it is generally advantageous to use basis sets which produce an as sparse representation of the Hamiltonian as possible, since this gives an efficient evaluation of matrix-vector products. In contrast, restricted active space calculations are not invariant under unitary transformations of the single-particle basis. Most 
obviously, this can be seen for a single determinant, where the choice of the orbitals is of primary importance. As it is well known, the optimal set of orbitals for a single determinant (in the sense of giving the lowest energy) is obtained by solving the Hartree-Fock equations. This demonstrates a principle that is generally valid in RAS calculations: the more the RAS space deviates from the Full CI space, the better the single-particle basis needs to be adapted. Therefore, to be confident, RAS calculations are usually carried out in a well adapted basis, regardless of the accuracy of the many-body space. Typical choices are the Hartree-Fock basis~\cite{Greenman_2010} or the basis obtained from a Multiconfigurational Hartree-Fock calculation. For both tasks, we employ our recently implemented Multiconfigurational time-dependent Hartree-Fock program~\cite{Hochstuhl_2011}.

\begin{figure}[!t]
 \begin{center}
   \includegraphics[width=\linewidth]{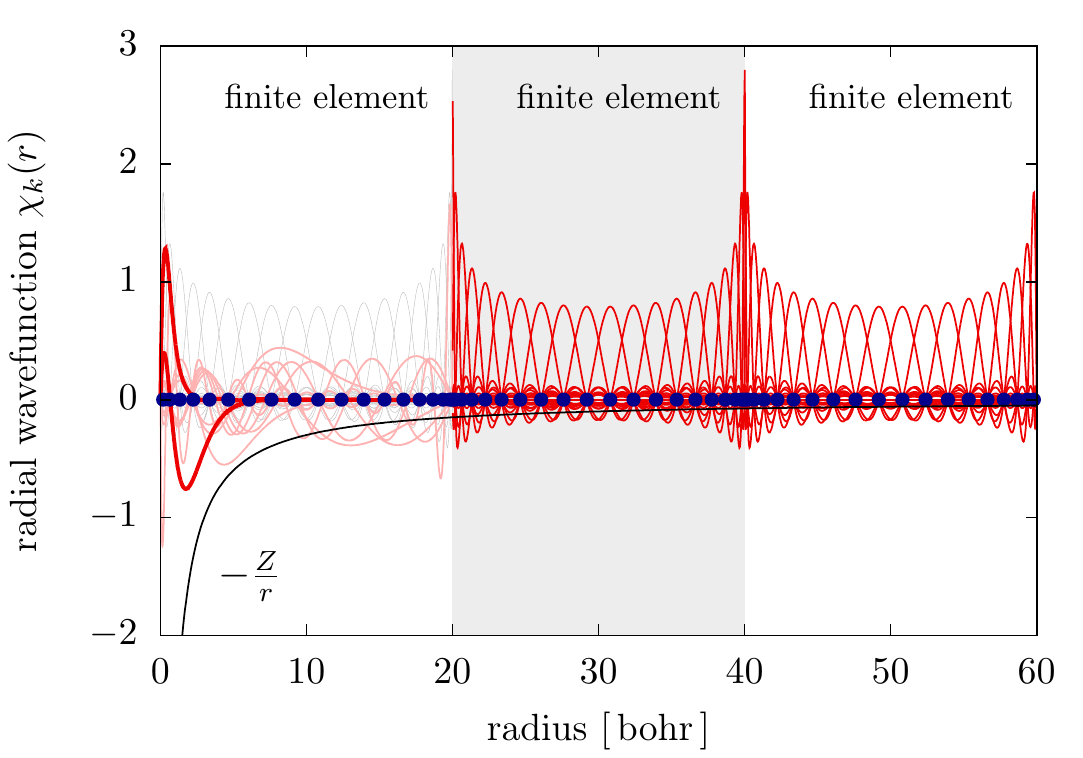}
 \end{center}
 \caption{\label{fig:mixed_basis}Radial part of the single-particle basis (\ref{eq:single_particle_basis}) used in this work, which consists of a combination of Hartree-Fock orbitals in the vicinity of the atom (left element), and FEDVR basisfunctions in the continuum (two elements on the right). In each element, the basisfunctions are constructed on a Gauss-Lobatto grid with 20 points. In the first element, the $1s$ and $2s$ orbitals are emphasized.}
\end{figure}

However, the usage of these adapted basis sets meets its limits in the treatment of photoionization. If the size of the basis $N_b$ becomes large, the creation and storage of the two-electron integrals [the $g_{pqrs}$ in Eq.~(\ref{eq:Hamiltonian_second_quant})] in the Hartree-Fock basis is a very tedious task, since both require at least an effort of $\mathcal O(N_b^4)$. Thus, the advantages of the grid-like FEDVR treatment [which roughly scales with $\mathcal O(N_b^2)$] were completely lost. To circumvent this, we apply a mixed basis, in which we perform the transformation onto the Hartree-Fock orbitals only in a small region $\mathcal B_1$ around the nucleus, while the remaining part $\mathcal B_2$ is described by the grid-like basis (for an illustration see again Fig.~\ref{fig:RASspatial}). By this, we obtain both a good description of the groundstate, which is assumed to lie almost entirely in $\mathcal H_1$, and of the continuum, and this at a moderate effort. For the construction of the mixed basis, we 
first solve the (Multiconfigurational) Hartree-Fock equations in $\mathcal B_1$ using the spherical FEDVR basis, which provides optimized occupied orbitals. The virtual orbitals, however, are usually delocalized and do not resemble excited atomic orbitals. Thus, to obtain appropriate pseudo-orbitals, we start from the ideal states and orthonormalize them to the bound states (for a more elaborate construction scheme, see \cite{Burke_1975}). As a result, we obtain the unitary transformation matrix $\mathbf U_1$, which transforms from the spherical FEDVR basis to the optimized orbitals. The matrix
\begin{align}
\mathbf U \ =\ \begin{pmatrix} \mathbf U_1 & 0 \\ 0 & \mathbf 1 \end{pmatrix}
\end{align}
is then used to transform the electron integrals in the total basis $\mathcal B$, Eq.~(\ref{eq:single_particle_basis_RAS}). The arising basis is illustrated in Fig.~(\ref{fig:mixed_basis}).
A similar basis consisting of a mixture of Gaussian-type and FEDVR orbitals has been used in Ref.~\cite{Yip_2010} to describe double ionization of Beryllium.

\section{Results}
In the present section, we apply the TD-RASCI method to the atoms Helium and Beryllium, for which we calculate the photoionization cross sections from an explicitly time-dependent treatment. The present results could alternatively be calculated using a variety of (mostly time-independent) methods like perturbation theory or Floquet theory, which are applicable to long pulse duration and yield sharp spectra. The following calculations can therefore be viewed as a proof-of-principle application, which constitutes a necessary first step for future application to explicitly time-dependent processes.
Our main objective is thereby to check whether the relevant states exist and are located at correct energy positions, so that they can participate in the simulations. Furthermore, we will not exploit the full computational power of the RAS scheme, but only use the two-fold division of space illustrated in Fig.~\ref{fig:RASspatial}, which is routinely applied in R-matrix calculations.
Thereby, we divide the coordinate space in two regions, one in the vicinity of the atom (region 1) and one outside (region 2). The boundary is placed at $R=20$ bohr, which appears to be a common choice also used in Ref.~\cite{Lysaght_2008}. In the outer region, we allow for a single electron, i.e. we investigate only single ionization, while the groundstate in the inner region is described using different approximation levels. 
The basic approximation is TD-CIS, where the wavefunction is described by one groundstate determinant plus all single-excitations. By including more determinants, we can then improve the description of the wavefunction in $\mathcal H_1$, and this way include relevant ionization channels.

\subsection{Helium}
The Helium atom is a thoroughly studied system for more than 40 years, and especially the last decade has seen an ever growing number of investigations, as, e.g., of pump-probe processes~\cite{Palacios_2009,Feist_2011}. Another topic of recent interest is the two-photon double ionization~\cite{Foumouo_2006,Feist_2008}, which has also been investigated in our former work using the Multiconfigurational time-dependent Hartree-Fock method~\cite{Hochstuhl_2011}. Here, we use the TD-RASCI method to consider single photoionization for photon energies in the direct and sequential regime of two-photon double ionization. In Helium, the idea to partition the single-particle basis has been applied several times before, see, e.g., Ref.~\cite{Venuti_1996} for a spatial partitioning, or Ref.~\cite{Feist_2008} for an optional partitioning in terms of angular momenta. Furthermore, R-matrix calculations have been performed on the single- and double ionization of Helium~\cite{Marchalant_1997,Nikolopoulos_2008,Guan_2009}.

\begin{figure}[!t]
 \begin{center}
   \includegraphics[width=1.0\linewidth]{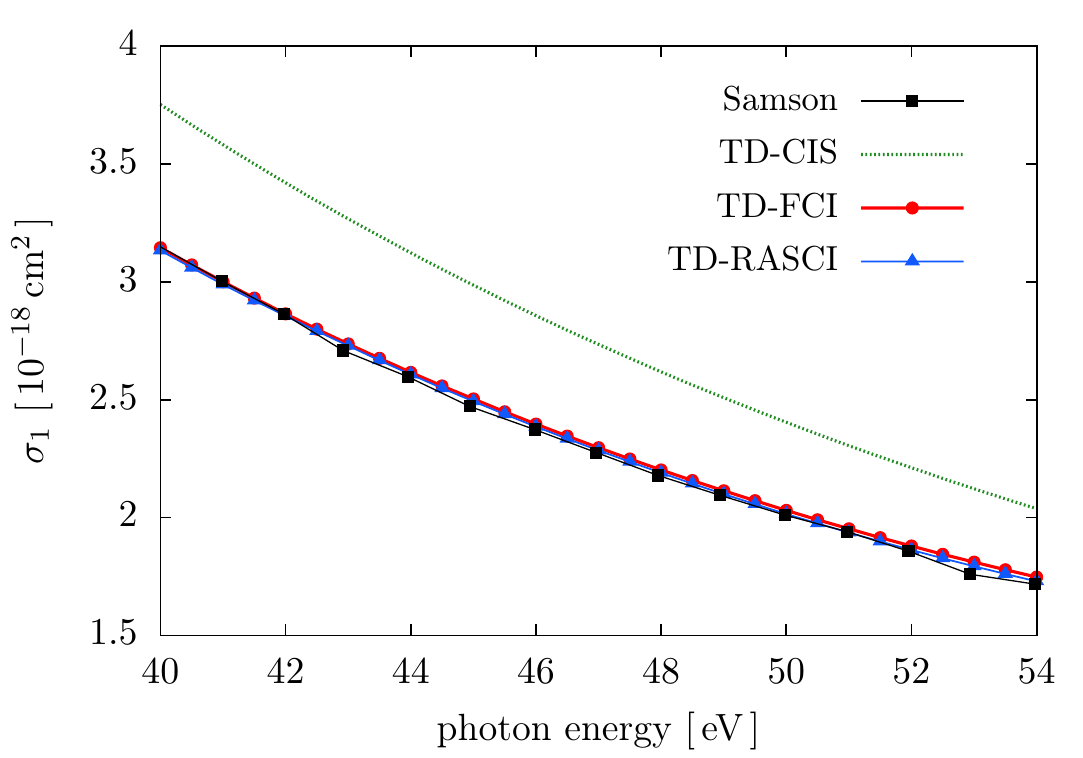}
 \end{center}
 \caption{\label{fig:helium_direct}(Color online) Single-ionization cross sections of Helium, calculated for a laser field with intensity $I=10^{12} W/cm^2$ and a squared-sine envelope of duration $T=400\,$a.u. The black dots show experimental results of Samson~\cite{Samson_1994}, red squares the TD-FCI reference results and blue triangles the TD-RASCI results, which all agree almost perfectly. The results of TD-CIS calculations are depicted by green triangles.}
\end{figure}

To obtain appropriate Helium orbitals, we solved the Hartree-Fock equations in the region $r\leq 20$ bohr using a radial basis consisting of $54$ FEDVR functions for the $1s$ orbital. For the virtual orbitals, we use single-hamiltonian eigenfunctions up to $f$-symmetry, with the ideal $ns$ orbitals orthonormalized to $1s$ HF-orbital. For the TD-RASCI method, we then employed a number of $10$ $s$-orbitals and $3$ $p$-orbitals to form the Full CI groundstate with vanishing spin- and angular-momentum projection values. A single electron can be excited out of this space either into the remaining ideal orbitals inside $r<20$ bohr or into the outside region, where a FEDVR basis is used with finite elements of length $4$ bohr containing $10$ basisfunctions each. The groundstate energy we obtain in this way is $E=-2.8894$ Hartree (the exact non-relativistic energy is $E=-2.9037$ Ha). We have further applied orbitals which have been optimized with the MCTDHF method. Though they yield an improved groundstate energy of 
$E=-2.9002$ Ha, the photoionization spectra were less accurate. This is probably due to a poorer representation of excited states, which is caused by the fact that the orbitals are optimized for the groundstate. On the groundstate, we apply a laser pulse with a squared-sine envelope of duration $T=400$ a.u. ($\sim 10$ fs) and intensity $10^{12}\, \text{W}/\text{cm}^2$. The ionization yields are calculated as the norm of the wavefunction in the region $r>20$ bohr and are extracted at the end of the pulse. The yields are related to the single-ionization cross sections through the formula given in Ref.~\cite{Foumouo_2006}.

the pulse. The yields are related to the single-ionization cross sections using the formula given in Ref.~\cite{Foumouo_2006}.

Figure~\ref{fig:helium_direct} depicts the cross section for photon energies in the direct regime of two-photon double ionization, i.e. $\omega=40$ eV to $\omega=54$ eV. Plotted are the results of TD-CIS, TD-RASCI, as well as experimental results of Samson \emph{et al.}~\cite{Samson_1994}. 
Further, we show time-dependent Full Configuration Interaction (TD-FCI) results taken from our former work~\cite{Hochstuhl_2011} (calculated for a field duration of $T=100$ a.u.).
All curves agree qualitatively and show a smooth, monotonically decreasing behavior. The TD-CIS curve, however, differs visibly from the experimental results, whereas the TD-FCI result, i.e. the exact solution of the two-particle problem, agrees well with the experiments. The TD-RASCI results matches the FCI reference result almost perfectly, which is remarkable as the calculations can be estimated to be about twenty times faster for the present parameters. One can thus conclude that, as expected, the largely reduced RAS Hilbert space suffices to describe the single-ionization, or, vice versa, that the excluded part of the Full CI space has a negligible occupation. The lower accuracy of the TD-CIS result, on the other hand, demonstrates the importance of an accurate description of the electronic structure.

\begin{figure}[!t]
 \begin{center}
   \includegraphics[width=1.0\linewidth]{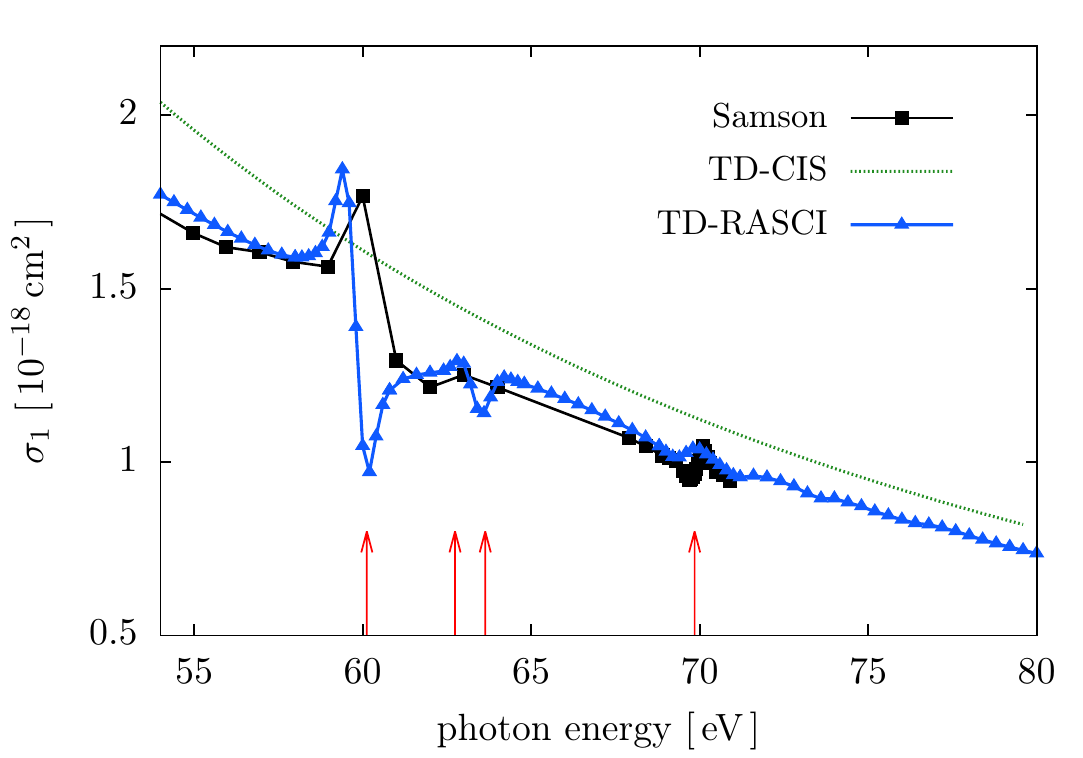}
 \end{center}
 \caption{\label{fig:helium_sequential}(Color online) Single-ionization cross sections of Helium for the same field parameters as in Fig.~\ref{fig:helium_direct}. Black squares are the experimental results of Samson \emph{et al.}~\cite{Samson_1994}, and red arrows mark the resonance positions of ${}^1P^o$-symmetry according to Scrinzi and Piraux~\cite{Scrinzi_1998}.}
\end{figure}

In Fig.~\ref{fig:helium_sequential}, we plot the cross section against photon energies in the sequential two-photon double-ionization regime, $\omega\geq54$ eV. At these photon energies, one observes several Fano resonances in the cross-section, which correspond to ionization channels where doubly-excited states become occupied and subsequently decay via autoionization occurs. The resonances arise due to the interference between these channels and the direct pathway, in which the electron is directly ionized into the continuum. The energies of the resonances with ${}^1P^o$-symmetry are indicated by red arrows, and are taken from a work of Scrinzi and Piraux who applied the complex-scaling method~\cite{Scrinzi_1998}. The experimental results of Samson \emph{et al.}~\cite{Samson_1994} are denoted by black squares. They clearly resolve the resonance at $60.1$ eV below the $N=2$ threshold, and the resonance at $69.9$ eV below the $N=3$ threshold (for a more detailed classification we refer to \cite{Domke_1991}). 
The TD-CIS results yield a smoothly decreasing curve that shows no peaks. This is not surprising as the doubly-excited bound states are missing in the expansion of the wavefunction, so that only the linear photoionization process is possible. In contrast, the TD-RASCI wavefunction includes the autoionizing states and thus contains the relevant ionization channels by construction. The TD-RASCI curve agrees well with the experimental results, and shows peaks at the correct resonance positions. Due to the restricted propagation time in our time-dependent method, the widths of the resonances are described less accurately, and much longer propagation times would be necessary to improve on that. For instance, the width of the $2s3p_{-}$ resonance requires a pulse of at least the same bandwidth, and hence a duration of at least $6$ ps~\footnote{An alternative to long propagation times is given in Ref.~\cite{Palacios_2008}, which involves the formal propagation to infinite time by solution of the driven equation}. 
We stress, however, that the focus of the present method is not on calculating accurate cross sections, but rather to allow for a time-dependent treatment of the many-body photoionization problem.

\begin{figure}
 \begin{center}
   \includegraphics[width=\linewidth]{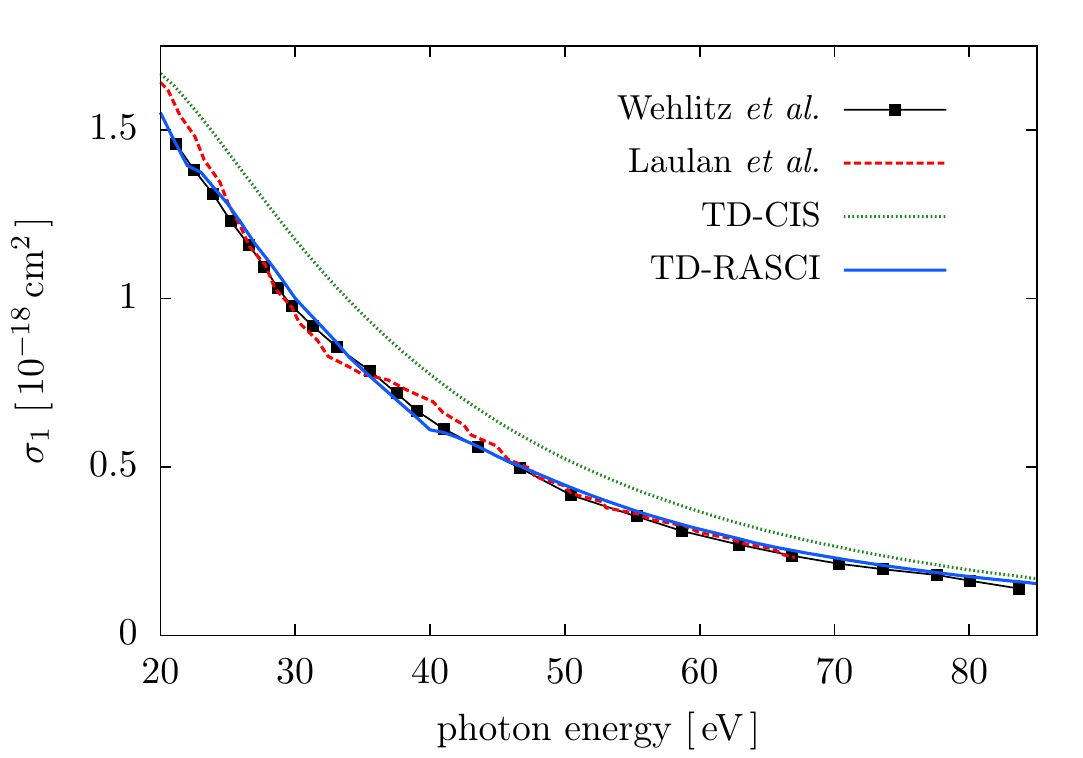}
 \end{center}
 \caption{\label{fig:be_cs_20_85}(Color online) Single-ionization cross section for Beryllium as a function of photon energy. Black squares denote the experimental results of Wehlitz~\emph{et al.}~\cite{Wehlitz_2002,Wehlitz_2005}, the dashed curve the time-dependent calculations of Laulan and Bachau~\cite{Laulan_2004}. Both are compared to TD-CIS (green) and TD-RASCI (blue) calculations.}
\end{figure}

\subsection{Beryllium}
Beryllium is the simplest atom with two closed shells. The ionization energy of its $2s$ orbital is experimentally found to be $9.32$ eV, and $123.35$ eV for ionization from the $1s$ orbital~\cite{Kramida_1997}. Therefore, when considering photon-energies below the $1s$-threshold, the large energetic separation of the two shells allows for a fixation of the $1s$ core-orbitals and thus for the reduction to an effectively two-particle problem. In this spirit, several works have been performed focusing on single- and double-electron photoionization. Most of them are following the time-independent approach, using, e.g., the R-matrix method~\cite{Kim_2000}, the relativistic random-phase approximation~\cite{Chi_1991}, hyperspherical calculations~\cite{Zhou_1995}, or the Multiconfigurational Hartree-Fock method~\cite{Saha_1989}. The time-dependent treatments are usually performed with the close-coupling approach~\cite{Colgan_2002,Laulan_2004} or a mixed-basis approach~\cite{Yip_2010}, both of which apply a 
reduction by freezing the $1s$ orbitals (or, occasionally, the $2s$ orbitals).

Here, for the TD-RASCI approach, we use a similar ansatz as before in Helium: after solving the Hartree-Fock equations and orthogonalizing the remaining ideal orbitals, we include $10$ $s$-orbitals and $2$ $p$-orbitals, which may be arbitrarily occupied by the $4$ electrons. The only restriction is that we allow only for single-excitations from the $1s$ orbitals, which is reasonable, as for the considered photon energy range and intensity double excitations from the $1s$ shell are negligible.
By this, we obtain a RASCI groundstate energy of $E=-14.618$ Ha, which has to be compared to the HF energy of $E=-14.573$ Ha and to the exact energy $E=-14.667$ Ha~\cite{Komasa_1995}.

\begin{figure}
 \begin{center}
   \includegraphics[width=\linewidth]{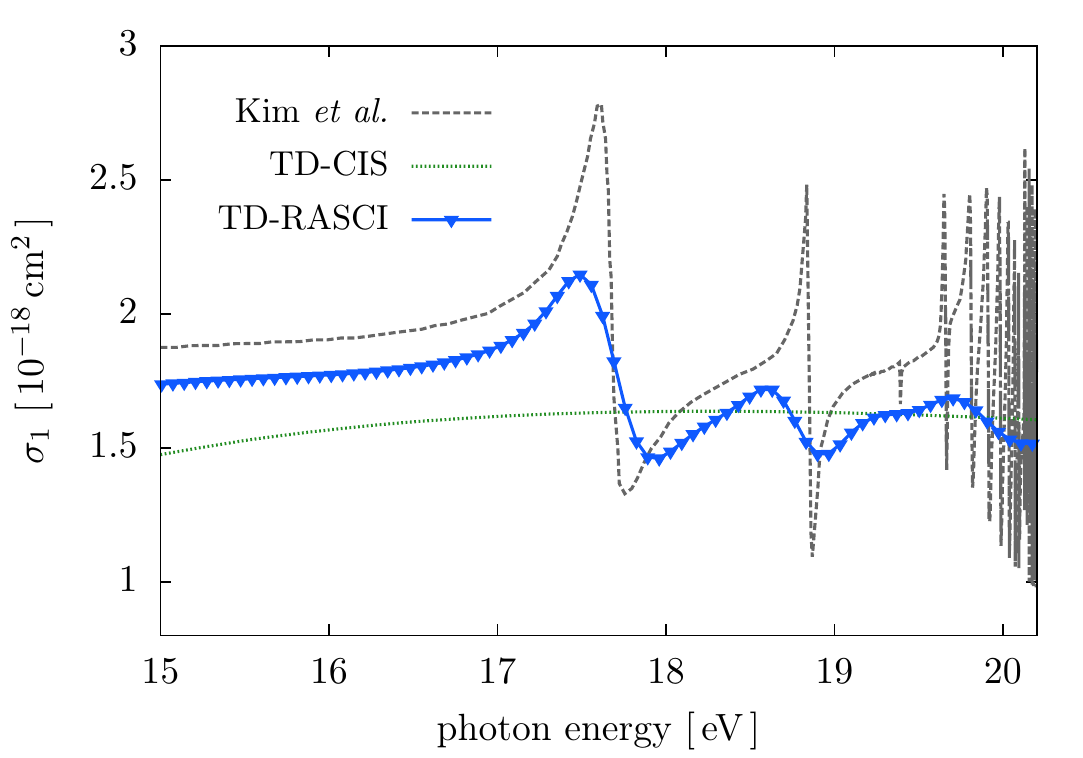}
 \end{center}
 \caption{\label{fig:be_cs_15_20}Single-photoionization cross section for Beryllium between the $\text{Be}^+(2p)$ and $\text{Be}^+(3s)$ thresholds. TD-CIS and TD-RASCI calculations are compared to the results of Kim~\emph{et al.}~\cite{Kim_2000}, which were obtained with the R-matrix method.}
\end{figure}

\begin{figure}
 \begin{center}
   \includegraphics[width=\linewidth]{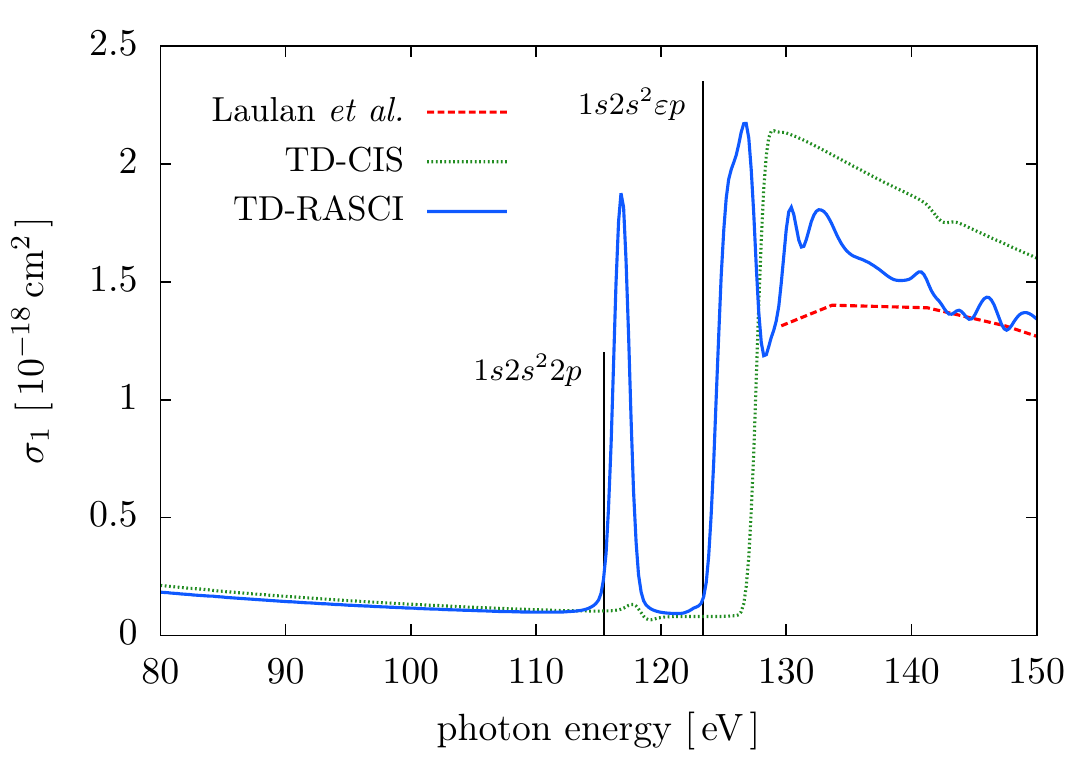}
 \end{center}
 \caption{\label{fig:be_cs_80_130}Photoionization cross-section of Beryllium around the $1s$ threshold. The red dashed line shows the results of Laulan and Bachau~\cite{Laulan_2004} obtained through a time-dependent method and fixation of the $2s$ orbitals, the green and blue curves our TD-RASCI and TD-CIS results. The black lines mark the resonance and ionization energy due to the NIST database.}
\end{figure}

Figure~\ref{fig:be_cs_20_85} shows the cross-sections for photon energies in the range $\omega=20$ eV to $85$ eV. The duration of the laser field is $400$ a.u. and its intensity is $10^{12}$ W$/\text{cm}^2$. The experimental results (black squares) are taken from two works of Wehlitz \emph{et al.}~\cite{Wehlitz_2002,Wehlitz_2005}. Further, we show the theoretical cross section of Laulan and Bachau, which was calculated with a time-dependent method and fixed $1s$ orbitals~\cite{Laulan_2004} as well. The TD-CIS results already show a good qualitative agreement. By improving the quality of the wavefunction, the TD-RASCI method is able to reproduce the experimental results almost perfectly. In Fig.~\ref{fig:be_cs_15_20}, we consider the resonances in between the $\text{Be}^+(2p)$ and $\text{Be}^+(3s)$ threshold, which were obtained from a pulse duration of $T=800$ a.u. The reference results are taken from an R-matrix study of Kim \emph{et al.}~\cite{Kim_2000}. One notices again, that not surprisingly TD-CIS is 
not 
adequate to model the peaks. The results are shown here only for reference. At the same time, the TD-RASCI method determines the gross structure correctly. Due to the limited bandwidth, our time-dependent approach does, however, not resolve the resonance region around $20$ eV. To obtain more accurate results, much longer propagation times and also a more accurate RASCI expansion allowing for double excitations to $d$-orbitals would be necessary. Finally, Fig.~\ref{fig:be_cs_80_130} concentrates on photon energies in between $\omega=80$ and $150$ eV, i.e. around the onset of the ionization of the $1s$ orbital at $123.35$ eV. Here, a pulse duration of $T=200$ a.u. is used. Again, we compare our results to those of Laulan and Bachau~\cite{Laulan_2004}, where, this time, the two $1s$ electrons have been considered as active and the $2s$ shell is held frozen. In contrast, the TD-RASCI approximation (and TD-CIS as a special case of it) treats ionization from both shells on the same footing, and is therefore able 
to adequately describe the transition region. In TD-CIS, the ionization of the $1s$ orbitals 
occurs at an energy of $\omega=128.72$ eV, which is the ionization energy obtained in Hartree-Fock approximation through Koopmans theorem. The more accurate TD-RASCI wavefunction is able to shift the onset energy towards the exact position. The same holds for the $1s\,2s^2\,2p$-resonance, which, unlike the resonances considered so far, is already contained in the TD-CIS approximation, as the state is obtained from the groundstate by a single excitation from $1s$ to $2p$. Again, the TD-RASCI approximation corrects the resonance position, and almost achieves the exact result. Above the $1s$ threshold, one encounters a large number of dense-lying resonances caused by the further excitation of the electrons from the $2s$ shell, which have been studied in detail by Voky \emph{et al.} using the time-independent R-matrix method~\cite{Voky_1992}. In contrast to the results of Laulan and Bachau, which show a smooth monotonically decreasing curve, the TD-RASCI method again reproduces the occurrence and position of the 
resonances which get occupied in the shake-up process.

\subsection{Pump-probe ionization of Beryllium}
As a final example we investigate the pump-probe process in Beryllium, which in contrast to the previous applications is accessible only through an explicitly time-dependent treatment. We consider the following scenario: Beryllium in its groundstate is ionized by an X-ray pulse with a photon energy of $200$ eV, an intensity of $10^{12}\,\text{W}/\text{cm}^2$ and a squared sine envelope with a duration of 20 cycles. At the same time, a single-cycle IR pulse with a wavelength of $780$ nm and an intensity of $10^{11}\,\text{W}/\text{cm}^2$ acts on the system. Following the usual streak camera principle, the delay $\delta$ between the X-ray and the IR pulse is varied and observables such as the ionization yields and momentum spectra are recorded as a function of $\delta$.

We use a model of the Beryllium atom, in which one electron can be ionized into the continuum. The ansatz to the wavefunction is given by
\begin{align}
\ket{\Psi(t)} \ = \ \sum_{\gamma} c_\gamma(t) \, \ket{\Psi_\gamma} + \sum_{\gamma,pq} c_{\gamma,pq}(t) \, \hat a^\dagger_p \hat a_q\ket{\Psi_{\gamma}}\,,
\end{align}
where the first term determines the included Slater determinants to model the groundstate and the second term collects all (unique) single-excitations of these determinants (note that this ansatz is a special case of the formalism in section \ref{sec:td_rasci}). We then use different levels of accuracy: in the basic approximation referred to as ($2s$) only the Hartree-Fock determinant and its single-excitations are included, i.e. it corresponds to a TD-CIS approximation. We further consider more accurate models in which we allow for double excitation up to the orbital $ns$ [denoted ($ns$)] and additionally up to the $mp$ orbital [denoted ($ns$,$mp$)]. As an example, ($5s,3p$) denotes a CISD wavefunction with double excitations allowed up to the $5s$ and $3p$ orbitals, and additionally all single excitations of these determinants. The presented calculations require in between half an hour on a single processor for the ($2s$) approximation and two days for the ($5s,3p$) approximation.

\begin{figure}[!t]
 \begin{center}
   \includegraphics[width=\linewidth]{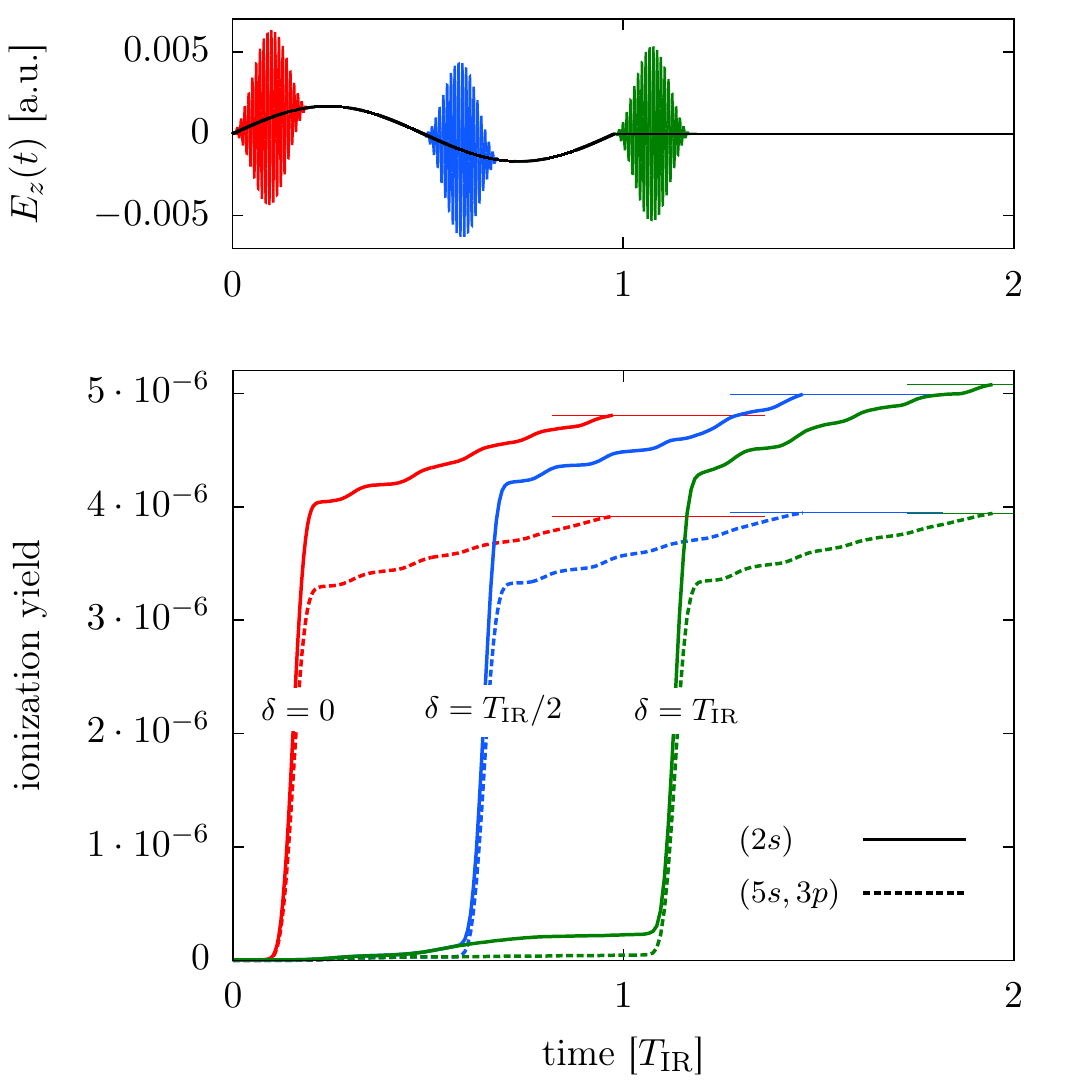}
 \end{center}
 \caption{\label{fig:single_ion}Total ionization yield of Beryllium subjected to the three different X-ray--IR pump-probe pulses shown on top. Shown are the results from a single-active electron calculation ($2s$) and the more accurate ($5s,3p$) TD-RASCI approximation (see text). For better comparison, the horizontal lines at the end of the pulses mark the total ionization yield.}
\end{figure}

Figure \ref{fig:single_ion} illustrates the time-dependent ionization yield for three different delays $\delta$ of the pump-probe pulse and the ($2s$) as well as the more sophisticated ($5s,3s$) approximation. As expected, all calculations show a steep rise of the yields during the action of the X-ray pulse. As suggested by Fig.~\ref{fig:be_cs_20_85}, the total yield is thereby larger for the single-active electron approximation. One further obtains relative differences for the three delays, which can be observed most obviously for the delay $\delta=T_{\text{IR}}$, where the X-ray field begins right after the IR pulse has passed. In the TD-CIS approximation ($2s$), at this time a significant portion has been ionized by the IR pulse, and accordingly the total yield is larger and also differs for the three delays. For the ($5s,3p$) approximation, the influence of the IR field on the ionization is much reduced and all three delays show a comparable ionization yield. This is caused by the fact that the more 
sophisticated TD-RASCI ansatz is able to model more accurately the polarization of the atom which is induced by the IR field.

\begin{figure}[!t]
 \begin{center}
   \includegraphics[width=\linewidth]{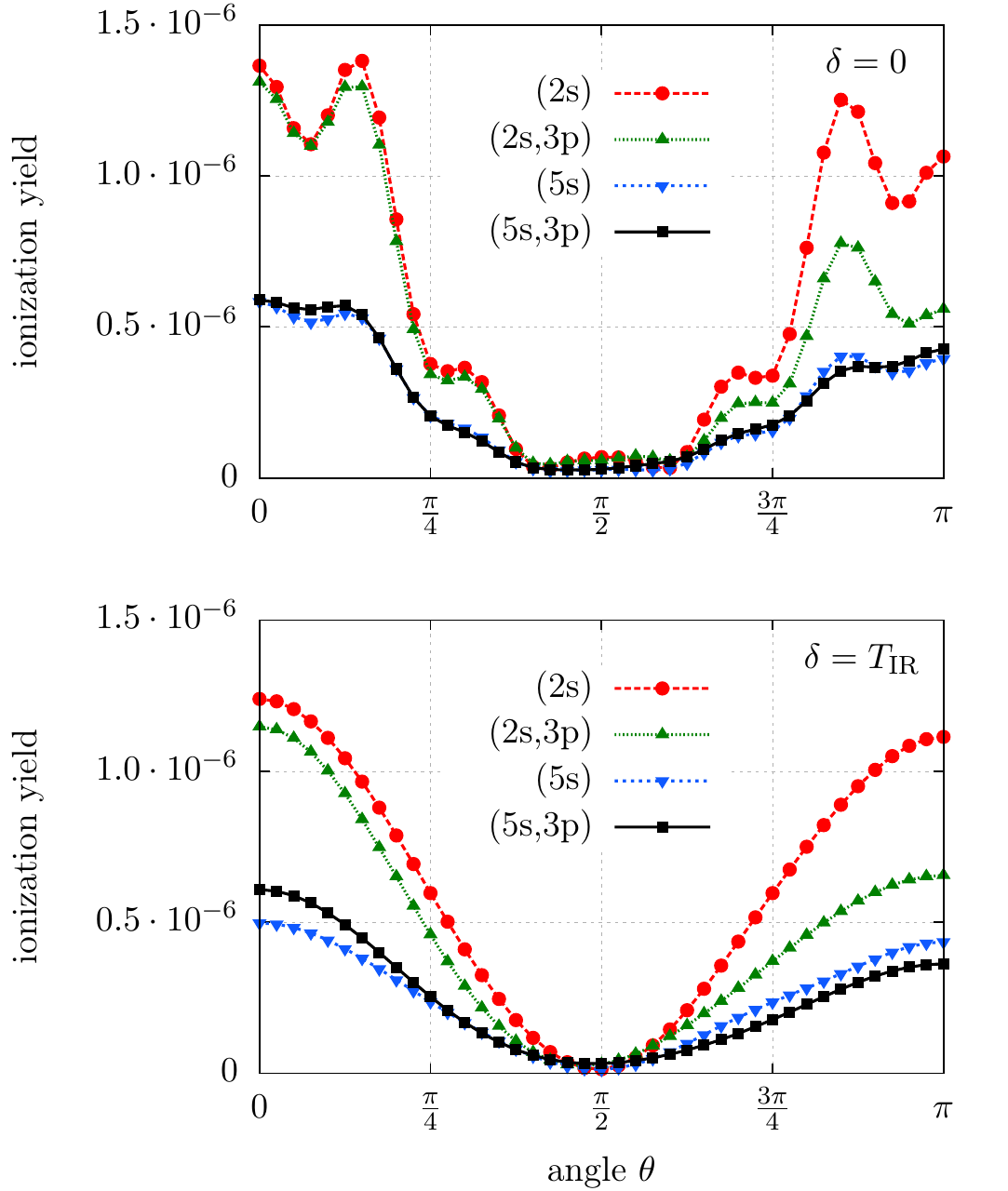}
 \end{center}
 \caption{\label{fig:angular_dist}Angular distribution of the photoionization yield of Beryllium subjected to the X-ray--IR pulse on top. Shown are the results of different RAS approximations (see text).}
\end{figure}

The angular distribution of the ionized part of the wavefunction is depicted in Fig.~\ref{fig:angular_dist}, for the two delays $\delta=0$ and $\delta=T_\text{IR}$. As before, the ionization yield is determined for each angle as the norm of the wavefunction outside a distance $r=20$ bohr from the nucleus. For the delay $\delta=0$, the photo-electron is excited to high angular-momentum states which cause the peculiar structure of the angular distribution.
This structure becomes largely damped upon inclusion of more $s$-orbitals into the expansion. Further, the use of more $p$-orbitals leads to a considerable asymmetry between forward ($\theta=0$) and backward ($\theta=\pi$) direction. For the delay $\delta=T_\text{IR}$, the X-ray pulse acts after the IR pulse has passed, which leads to ionization dominantly to lower partial waves and therefore to a smoother distribution. Nevertheless, one observes a noticeable influence on the ionization yields. The best TD-RASCI approximation, ($5s,3p$), again gives an ionization yield which is reduced by a factor of roughly $2.5$ as compared to the TD-CIS curve and, further, predicts a larger ionization in forward direction. This is remarkable as there is no IR field acting on the atom, and one could also expect a symmetric distribution as it is obtained in the normal photoionization. Hence, it appears that the IR field induces electronic motion which remains after the pulse has passed and affects the photoionization 
process. The detailed investigation of the characteristics of this special photoionization process will be part of a future work.

\section{Conclusion}
In this work, we introduced the time-dependent restricted active space Configuration Interaction (TD-RASCI) method for the ab-initio simulation of photoionization processes. Though well known in quantum chemistry and based on a conceptually simple idea, to our knowledge the TD-RASCI method has in its full generality not been applied to solutions of the time-dependent Schrödinger equation so far. Only specialized variants have been employed, like the single- and two-active electron approximation or time-dependent Configuration Interaction singles. In particular, the TD-RASCI method bears several similarities to the time-dependent R-matrix method, which can be considered the most successful ab-initio approach to many-body atoms at present. However, though the underlying idea is nearly identical, the actual implementation differs considerably and is, in our opinion, conceptually much simpler for the TD-RASCI scheme. In fact, the only task is the selection of an appropriate active space, for which we presented a 
systematic strategy. This accomplished, the derivation of the equations of motion proceeds  as in the Full CI method, and is as easy as going from Eq.~(\ref{eq:wavefunction_expansion}) to Eq.~(\ref{eq:Schroedinger_equation_discretized}): all one has to do is to insert the restricted ansatz into the Schrödinger equation, and arrive at a matrix equation for the expansion coefficients. We further presented the specializations required to efficiently treat photoionization processes, namely the use of a spherical FEDVR single-particle basis and, in order to obtain an appropriate initial state, the transformation onto a mixed basis set.

The TD-RASCI method has been applied to the calculation of single-ionization cross sections of Helium and Beryllium as well as to an X-ray--IR pump-probe scenario in Beryllium, and we particularly examined the quality of different RAS approximations. It was shown that time-dependent Configuration Interaction singles (TD-CIS) method is able to qualitatively describe the experimentally measured cross-sections in the linear photoionization regimes. However, by definition, it lacks the description of resonances, simply because the relevant doubly-excited states are not included in the wavefunction. By taking them into account in the TD-RASCI method, we are able to appropriately resolve the resonances. Due to the time-dependent approach, the energy-resolution of the spectra is less accurate than in time-independent calculations. Yet, this is of no practical significance in time-dependent applications, where it only matters that the relevant states exist and are located at the correct energy positions. This 
requirement being fulfilled, we are confident that the TD-RASCI method is well suited for studying a large class of essentially time-dependent physical processes, including pump-probe scenarios or the Auger decay of many-electron systems. Our future work will therefore concentrate on an application of the TD-RASCI method to these problems.

\section{Acknowledgements}
We gratefully acknoledge support by the Bundesministerium für Bildung und Forschung via the project FLASH and by a CPU grant at the North German Supercomputer Center (HLRN, grant shp0006).


%

\end{document}